\pdfoutput=1

\documentclass[11pt]{article}

\usepackage[final]{acl}

\usepackage{times}
\usepackage{latexsym}
\usepackage[T1]{fontenc}
\usepackage[utf8]{inputenc}
\usepackage{microtype}
\usepackage{inconsolata}
\usepackage{numprint}

\usepackage{array}

\usepackage[TABBOTCAP]{subfigure}
\usepackage{amssymb}
\usepackage[shortlabels]{enumitem}
\usepackage{placeins}
\usepackage{bbold}
\usepackage{tikz-dependency}
\usepackage{algorithm}
\usepackage{algpseudocode}
\usepackage{multirow}
\usepackage{booktabs}

\usepackage{color}
\usepackage{helvet}
\usepackage{textcomp}
\usepackage{booktabs}

\usepackage{color}
\usepackage{bbm}
\usepackage{graphicx}
\graphicspath{ {images/} }
\usepackage{amsmath}
\usepackage{amsfonts}

\usepackage{hyperref}
\usepackage{url}
\usepackage{tabularx}
\usepackage{tcolorbox}

\usepackage{xcolor}
\usepackage{color-edits}
\usepackage{array}
\addauthor{sw}{orange}
\usepackage{XX-slnotation}
\renewcommand{\slReals}{\mathbb{R}}

\renewcommand{\mlOutput}{\hat{y}}

\newcommand{\query}{q}

\newcommand{\rModel}{\mathcal{R}}

\newcommand{\MoRModel}{\mlPredictor}

\newcommand{\doc}{d}
\newcommand{\corpus}{\mathcal{D}}

\newcommand{\retrievalScores}{\mathbf{s}}

\usepackage{xspace}
\newcommand{\mor}{\texttt{MoR}\xspace}

\usepackage{enumitem}
\newlist{inlinelist}{enumerate*}{1}
\setlist*[inlinelist,1]{%
  label=(\roman*),
}

\title{MoR: Better Handling Diverse Queries with a Mixture \\ of Sparse, Dense, and Human Retrievers}
\author{
 \textbf{Jushaan Kalra$^{\dagger1}$}\thanks{\quad $^{\dagger}$ denotes equal contribution. Corresponding contact email addresses: \{xinranz3,sherryw\}@andrew.cmu.edu. Our code is available at  \url{https://github.com/Josh1108/MixtureRetrievers}.}\quad
 \textbf{Xinran Zhao$^{\dagger1}$}\quad
 \textbf{To Eun Kim$^1$}\quad
 \textbf{Fengyu Cai$^2$}\quad \\
 \textbf{Fernando Diaz$^1$}\quad
  \textbf{Tongshuang Wu$^1$}\\
 $^1$Carnegie Mellon University, $^2$Technical University of Darmstadt
}

\date{}

\begin{document}
\maketitle
\begin{abstract}

Retrieval-augmented Generation (RAG) is powerful, but its effectiveness hinges on which retrievers we use and how. Different retrievers offer distinct, often complementary signals: BM25 captures lexical matches; dense retrievers, semantic similarity. Yet in practice, we typically fix a single retriever based on heuristics, which fails to generalize across diverse information needs.
Can we dynamically select and integrate multiple retrievers for each individual query, without the need for manual selection?
In our work, we validate this intuition with quantitative analysis and introduce a \textit{mixture of retrievers}: a zero-shot, weighted combination of heterogeneous retrievers.
Extensive experiments show that such mixtures are effective and efficient: 
Despite totaling just 0.8B parameters, this mixture outperforms every individual retriever and even larger 7B models—by +10.8\% and +3.9\% on average, respectively.
Further analysis also shows that this mixture framework can help incorporate specialized non-oracle \textit{human} information sources as retrievers to achieve good collaboration, with a 58.9\% relative performance improvement over simulated humans alone.

%we validate such intuition with quantitative analysis and propose to leverage it to build a \textit{mixture of retrievers} that utilizes weighted outputs of different retrievers in a zero-shot manner. Extensive experiments show that such a mixture, with 0.8 billion parameters cumulatively, can work better than every single retriever as well as larger retrievers with 7 billion parameters, with on average +10.8\% and +3.9\% compared to the best component and large retrievers, respectively. 
%Further analysis shows how this mixture framework can help incorporate specialized non-oracle \textit{human} information sources as retrievers to achieve good collaboration, with a 58.9\% relative performance improvement over simulated humans alone.

% }\fcc{may consider highlighting the significant size-wise difference.}
\end{abstract}

\section{Introduction}\label{sec:introduction}

% \comment{we provide higher-level intuitions and conceptual descriptions here and move the details into the next section}
% p1: RAG; and the problem of retriever selection
Although Retrieval Augmented Generation (RAG) \citep{lewis2020retrieval} has been shown to improve the reliability and reduce the hallucination of large language models (LLMs), 
no single retriever is optimal for all queries.
For example, in encyclopedic question answering tasks such as Natural Questions~\cite{kwiatkowski-etal-2019-natural}, embedding-based retrievers like DPR~\cite{DBLP:conf/emnlp/KarpukhinOMLWEC20} often outperform
token-based approaches like BM25~\citep{DBLP:journals/ftir/RobertsonZ09}. In contrast, in specialized domains such as medicine and biology, token-based approaches remain a strong baselines~\cite{thakur2021beir}. 
Because retrieval effectiveness can vary significantly across domains and even across queries, finding a universally optimal retriever remains an open problem,  highlighting the importance of understanding query-level comparative advantages among retrievers, rather than relying solely on aggregate performance.
As a result, to accommodate real-world applications with diverse query types \cite{blendedRAG}, a core challenge is to select appropriate retrievers and combine their content.

\begin{figure*}[!t]
\vspace{-0.2in}
    \centering
    \hyperref[sec:method]{%
    \includegraphics[clip,trim={2cm 25cm 22cm 0.2cm},width=0.95\linewidth]{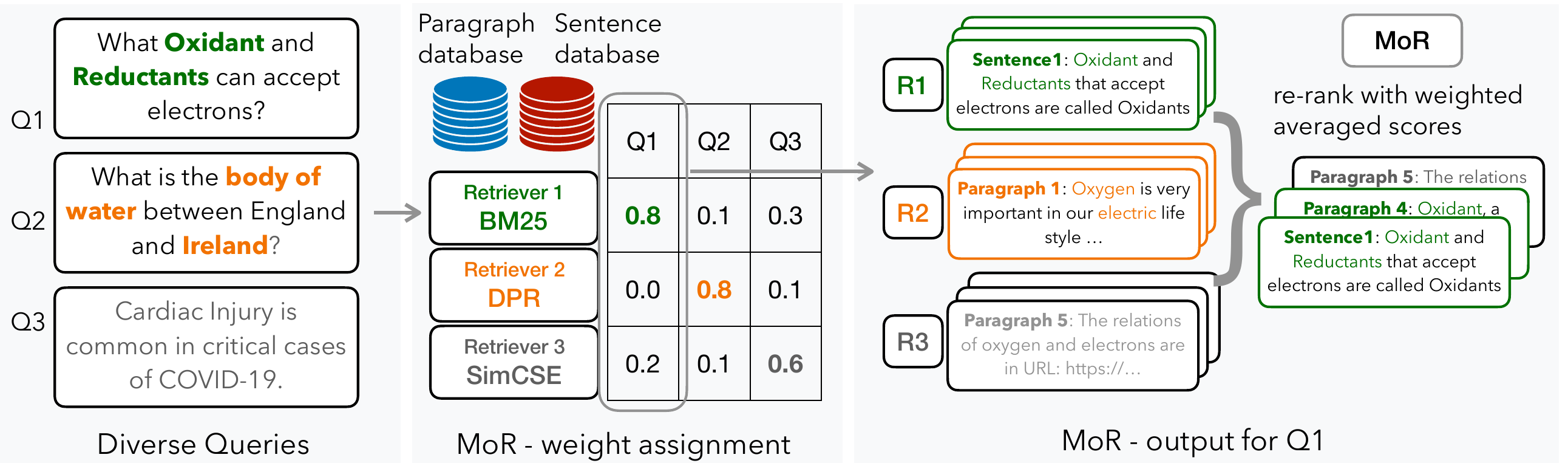}
    }
    % \vspace{-9pt}
\caption{We demonstrate Mixture of Retrievers (\mor), which handles diverse queries—such as keyword-based, general, and factual questions—by combining multiple retrievers: BM25, DPR, and SimCSE. \mor assigns trustworthiness weights to each retriever in a zero-shot manner, then aggregates and re-ranks document scores across databases of varying granularity to produce the final output.}
    \label{fig:motivation}
    \vspace{-0.1in}
\end{figure*}

% p2 To address this challenge we ask: \textit{Can this burden be mitigated through automatic selection and aggregation of retrievers?}
In this work, we propose a solution in the form of \textbf{Mixture-of-Retrievers} (\mor) framework. Inspired by the Mixture-of-Experts architecture~\cite{jacobs:moe,shazeer2017}, \mor dynamically selects and combines retrievers for each query by leveraging signals collected both before and after retrieval, eliminating the need for manual retriever selection.
% we collect automatic signals before and after conducting retrieval to decide what external knowledge to fetch based on the weighted average of the outputs from a selected set of retrievers. 
Specifically, as shown in Figure~\ref{fig:motivation}, \mor adopts a multi-granularity retrieval strategy~\cite{chen2023dense} to expand the pool of retrievers and exploit their complementary strengths operating on different semantic units. 
% \fcc{There might not be logical relations between mixed-granularity and different retriever types. Therefore, the word \textit{their} will be ambiguous, between retrievers or between mixed-granularity and mixed retrievers?}
Based on work from aggregated search \cite{diaz:vertical-chapter}, we consider retriever-trustworthiness signals at different stages:
\begin{inlinelist}
\item pre-retrieval signals, where we extend the notion of model familiarity from the \textit{when-to-retrieve} literature~ \cite{mallen-etal-2023-trust, zhao2023thrust} to the retriever-level, and  
\item post-retrieval signals, where we design signals akin to query performance prediction~\cite{10.1145/1277741.1277841,10.5555/2600139,10.1145/3077136.3080665,10.1145/3673791.3698438}. 
\end{inlinelist}
Using these signals, we compute per-query, per-retriever weights subsequently used to adjust the relevance scores, enabling effective re-ranking of the aggregated retrieval results from the entire pool of retrievers.
% With the weight per query per retriever, we utilize a parametric combination to comprehensively leverage query-document relevance scores from all retrievers.

% \fcc{FC: Figure 1 might be clickable.}

We conduct extensive experiments on four complex scientific domain retrieval tasks featuring diverse queries~\cite{boteva2016full, cohan-etal-2020-specter, wadden-etal-2020-fact, welbl-etal-2017-crowdsourcing}. Our results show that variants of \mor outperform individual retrievers, including the state-of-the-art GritLM~\citep{muennighoff2024generative} in both retrieval and RAG tasks. 
This shows that \mor determines the strengths and weaknesses of different retrievers and effectively automates retriever selection.

Notably, our further analysis demonstrates that the mixture and fusion strategy generalizes across a wide range of retriever types, including token-based, embedding-based, and even ``human retrievers", where humans provide noisy but useful information in response to queries. This highlights the potential of human-LLM collaboration in knowledge-intensive tasks, achieving on average 1.4$\times$ performance of \mor alone and a +58.9\% relative gain over purely simulated human responses.

% Further analysis shows how \mor can potentially help incorporate noisy human information sources to get further performance gain and how the efficiency can be improved with thresholding.
% \swcomment{Right now the human part feels a bit out of place; Maybe just merge it into prior descriptions saying "can automatically select retrievers and figuring out the strengths and weaknesses of different retrievers; Novelly, through further analysis, we show that the idea of mixture and fusion generalizes to different kinds of retrievers, including textual-based, embedding-based, and even ``human retrievers'' where humans provide LLMs with noisy information in response to LLM inquiries, which also points to the promise of human-LLM collaboration on knowledge tasks."}

In summary, our main contributions are:

\begin{itemize}[labelwidth=*,leftmargin=1.3em,align=left, topsep=0pt, nosep]
\item We identify the query-level comparative advantages and propose the \mor architecture to improve retrieval performance with \textit{folk wisdom}.

\item We design a zero-shot method to construct \mor with multi-granularity deep fusion, pre-retrieval signals, and post-retrieval performance prediction, which achieves improved performance and robustness over existing retrievers.

\item Further analysis on simulated noisy human experts shows \mor can potentially help estimate human retriever trustworthiness from a give corpus, and improve the human-LLM collaboration on knowledge-intensive tasks.
\end{itemize}
\section{Related Work}\label{sec:related_work}
\textbf{Adaptive treatment in RAG.}
% Retrieval-Augmented Generation \cite{lewis2020retrieval}, a prominent instance of Retrieval-Enhanced Machine Learning \cite{kim2024reml}, has been widely adopted to enhance the factuality and grounding of generated outputs. 
Recent advancements in RAG have increasingly adopted to apply diverse strategies to enhance adaptability across diverse tasks. \citet{blendedRAG} seek to combine lexical and semantic retrievers via designing fixed query types. In parallel, various studies have investigated adaptive query treatment mechanisms for differen purposes, including identifying relevant knowledge sources~\cite{Guerraoui25FederatedRAG, lee2024routerretriever}, adapting retrieval strategies based on query complexity~\cite{mallen-etal-2023-trust, zhao2023thrust,jeong-etal-2024-adaptiveRAG}, optimizing retrieval cost-effectiveness \cite{mu-etal-2024-query}, supporting multimodal tasks \cite{yeo2025universalrag} and ranking retrievers through learning~\cite{kim2025ltrrlearningrankretrievers}. Instead of choosing fixed specific retrievers, our work dynamically utilizes the information from all retrievers to combine their complementary advantages.
% \fcc{Our work ...}

\noindent\textbf{Distributed Information Retrieval.}
Advances in the adaptive RAG systems are deeply rooted in foundational research in information retrieval (IR), particularly the field of distributed information retrieval. Concepts such as resource selection \cite{dai2017ltrResources, 10.1145/3624918.3625330}, distributed and federated search \cite{callan1995searching, callan2002distributed, diaz2010federatedAggregated}, meta-search \cite{glover1999metasearch, chen2001metaspider}, and query performance prediction \cite{10.1145/1277741.1277841,10.5555/2600139,10.1145/3077136.3080665,10.1145/3539618.3592082,10.1145/3673791.3698438} for determining where to search are directly relevant to the modern multi-retriever set up in RAG systems. Similarly, ideas from aggregated search \cite{arguello2017aggregated}, and rank fusion techniques \cite{cormack2009reciprocal} have informed how retrieval signals should be combined across multiple sources, highlighting the deep connections between routing in RAG architectures and traditional IR research. In our work, we study how the semantics of dense retriever representation space serve as a strong zero-shot signal to combine information from multiple sources.

% \section{Is there a folk wisdom among retrievers?}
% \fcc{May use a shorter title to make it in one line? e.g., Any folk wisdom among retrievers?}
\section{Any folk wisdom among retrievers?}
%We first quantitatively validate our assumption on the \textit{folk wisdom} of retrievers, where similar techniques have been widely explored in various other pipelines in the machine learning community
Following existing practice~\cite{Ngo_2022,wang2023selfconsistencyimproveschainthought}, we first quantitatively validate our assumption that retrievers have varying strengths, and establish the potential of mixturing through simple query routing.

\begin{table*}[t]
\centering
\resizebox{\textwidth}{!}{

\begin{tabular}{r|l |l| l | p{7cm} | p{7.5cm}}
\toprule
\textbf{Retriever} & \textbf{Type} & \textbf{Backbone} & \textbf{Params} & \textbf{Training Signal} & \textbf{Strength} \\
\midrule
\multicolumn{6}{c}{\emph{Retrievers relying on general-purpose embeddings}}\\
\midrule
BM25%\newline\citep{10.1145/2682862.2682863} 
    & Sparse & TF-IDF & N/A & None (rule-based) & Strong zero-shot; fast and interpretable \\

SimCSE%\newline\citep{gao-etal-2021-simcse} 
    & Dense & BERT-base & 110M & Self-supervised contrastive on Wikipedia & Simple training; effective sentence encoder \\

Contriever%\newline\citep{izacard2022unsupervised} 
    & Dense & BERT-base & 110M & Unsupervised contrastive (web + Wikipedia) & No labels needed; strong zero-shot \\
    
\midrule
\multicolumn{6}{c}{\emph{Retrievers trained on query-document pairs}}\\
\midrule

DPR%\newline\citep{karpukhin-etal-2020-dense} 
    & Dense & Dual BERT-base & 110M & Supervised on QA (e.g., SQuAD) & Strong in-domain accuracy; widely adopted \\

ANCE%\newline\citep{xiong2021approximate} 
    & Dense & Dual BERT-base &110M & Contrastive with ANN negatives & Hard negative mining improves retrieval quality \\

TAS-B%\newline\citep{hofstatter2021efficiently} 
    & Dense & BERT-base & 66M & Distilled from ColBERT & Efficient; distilled from interaction-heavy model \\

GTR%\newline\citep{ni-etal-2022-large} 
    & Dense & T5-base & 220M & QA pretrain + MS MARCO finetune & Generalizes well \\

MPNet%\newline\citep{song2020mpnet} 
    & Dense & MPNet & 110M & Permutation + position-aware pretraining & Strong encoder for semantic similarity \\

\midrule
\multicolumn{6}{c}{\emph{Retrievers based on large language models}}\\
\midrule
RepLLaMA%\newline\citep{ma2023finetuningllamamultistagetext} 
    & Dense & LLaMA-7B
    & 7B & Supervised DPR-style dual-encoder & High quality; strong long-context handling \\

GritLM%\newline\citep{muennighoff2024generative} 
    & Dense & 7B LM & 7B & Joint generative + embedding training & SOTA on MTEB~\citep{muennighoff2022mteb}; \newline dual-purpose model  \\ 
\bottomrule
\end{tabular}
}
\vspace{-0.1in}
\caption{Comparison of retrievers used in or compared with our mixture. We include both sparse and dense models, varying in architecture, size, and training signals. Additional details are in Section~\ref{sec:appendix_retriever_details}.}
\label{tab:retriever_comparison}
\vspace{-0.1in}
\end{table*}

\subsection{Preliminary}
\label{sec:experiment_settings}

\paragraph{The task of retrieval.} We consider a set of retrievers $L_\rModel$, where each retriever $\rModel_i \in L_\rModel$. Then the task of a single retriever is: given a query $\query$ and a corpus $\corpus = \{\doc_j\}$, each retriever $\rModel_i$ assigns a relevance score to each document $\doc_j$ using its own scoring function $s_{i}(\query, \doc_j)$ (e.g., cosine similarity). This results in a score vector $\retrievalScores_i \in \slReals^{|\corpus|}$, where $|\corpus|$ is the size of the corpus. These scores can then be used to rank the documents.

\paragraph{Datasets and metrics.} Then we build a test bed for various retrievers on retrieval tasks with diverse query types, i.e., compared to encyclopedic questions in Natural Questions~\cite{kwiatkowski-etal-2019-natural}, these tasks involve complex query-document relations, e.g., there are multiple conditions organized in a first-order logic~\cite{cai-etal-2024-mixgr}. Specifically, with similar setting as BEIR~\cite{thakur2021beir}, we consider:
\begin{inlinelist}
\item NFCorpus~\cite{boteva2016full}: a medical retrieval dataset with non technical natural language queries and a complex terminology-heavy corpus with medical documents;
\item SciDocs~\cite{cohan-etal-2020-specter}: a scientific document retrieval task with diverse subtasks such as citation prediction, paper recommendation, etc, with a corpus of documents from more than ten domains including art, business, computer science, geology, etc;
\item  SciFact~\cite{wadden-etal-2020-fact}: a retrieval task with expert-written scientific claims and evidential abstracts as the corpus; and 
\item SciQ~\cite{welbl-etal-2017-crowdsourcing}: a large retrieval dataset with domain-specific text and science exam style questions.
\end{inlinelist}
% (1) NFCorpus~\cite{boteva2016full}: a medical retrieval dataset with non technical natural language queries and a complex terminology-heavy corpus with medical documents; (2) SciDocs~\cite{cohan-etal-2020-specter}: a scientific document retrieval task with diverse subtasks such as citation prediction, paper recommendation, etc, with a corpus of documents from more than ten domains including art, business, computer science, geology, etc; (3) SciFact~\cite{wadden-etal-2020-fact}: a retrieval task with expert-written scientific claims and evidential abstracts as the corpus; (4) SciQ~\cite{welbl-etal-2017-crowdsourcing}: a large retrieval dataset with domain-specific text and science exam style questions.
Further statistics and details are shown in Table~\ref{tab:dataset_stats} in the appendix. 
Following~\cite{cai-etal-2024-mixgr}, we use Normalized Discounted Cumulative Gain (NDCG@K) to compare the retrieval performance, where K denotes the number of top retrieved documents considered.

\paragraph{A diverse set of retrievers.}

% \swcomment{and, is it easier to create a table that summarizes the key characteristics, and put this full list into appendix? my chat with chatgpt \url{https://chatgpt.com/share/682949b8-7e04-800a-8497-01d986fa2344} - and see table 1. seems can generally convey the idea that different retrievers have different strength}

We consider both sparse and various BERT-sized~\cite{devlin-etal-2019-bert} dense retrievers as candidates of $L_\rModel$ to create our mixture of retrievers, including (1) unsupervised: BM25 \citep{10.1145/2682862.2682863}, SimCSE \citep{gao-etal-2021-simcse}, and Contriever \citep{izacard2022unsupervised}; (2) supervised: DPR \citep{karpukhin-etal-2020-dense},  ANCE \citep{xiong2021approximate}, TAS-B \citep{hofstatter2021efficiently}, GTR \citep{ni-etal-2022-large}, and MPNet~\cite{song2020mpnet}. The cumulative number of parameters of $L_\rModel$ is 836 million, i.e., 0.836 B. In addition, we also set a competitive performance reference with various large-language-model-based retrievers with 7 billion (7B) parameters:  RepLLaMA~\citep{ma2023finetuningllamamultistagetext} and GritLM~\citep{muennighoff2024generative}.

As depicted in Table~\ref{tab:retriever_comparison}, the above retrievers constitute a diverse set that varies in parameter sizes, backbone architectures, and training signals.  We include further detailed textual descriptions and model checkpoints we used in Section~\ref{sec:appendix_retriever_details}.

\subsection{Validating the folk wisdom}
\label{sec:oracle_retrievers}

% \begin{table}[t]
%   \centering
%   \resizebox{0.48\textwidth}{!}{%
%   \begin{tabular}{lccccc}
%     \toprule
%     Win (\%)& ANCE & Contriever & DPR & SimCSE & TAS-B \\
%     \midrule
%     ANCE & -- & 11.7 / 52.7 & 33.6 / 9.2 & 38.5 / 20.1 & 14.1 / 41.7 \\
%     Contriever & 52.7 / 11.7 & -- & 59.0 / 8.8 & 66.4 / 5.3 & 33.2 / 15.5 \\
%     DPR & 9.2 / 33.6 & 8.8 / 59.0 & -- & 32.9 / 26.9 & 11.7 / 48.8 \\
%     SimCSE & 20.1 / 38.5 & 5.3 / 66.4 & 26.9 / 32.9 & -- & 8.5 / 51.6 \\
%     TAS-B & 41.7 / 14.1 & 15.5 / 33.2 & 48.8 / 11.7 & 51.6 / 8.5 & -- \\
%     \bottomrule
%   \end{tabular}
%   }%
%   \caption{Comparison table of different models (\xrc{this will be convert into figures eventually.})}
%   \label{tab:prelim_1}
% \end{table}

\begin{figure}[t]
    %\vspace{-0.2in}
    \centering
    \includegraphics[clip,trim={0.2cm 0.1cm 0.2cm 0.2cm},width=\linewidth]{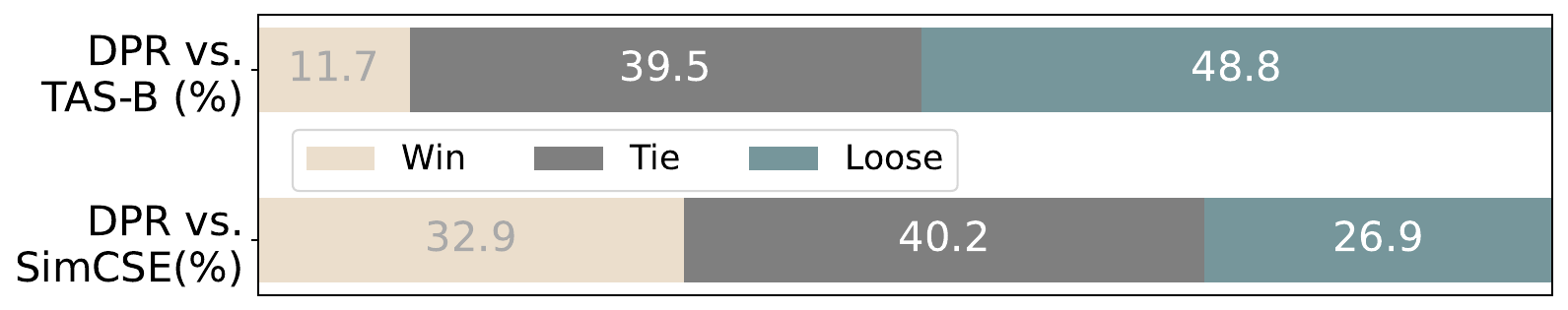}
    % \vspace{-0.1in}
\caption{{Performance comparison of different retrievers on SciFact. \textit{A Wins B} denotes that the gold entry appears in the top 20 retrieved documents of A, not B. The rates denote the micro-average of all queries, which shows clear comparative advantages.}}
\label{fig:prelim_1}
% \vspace{-0.2in}
\end{figure}

% \swcomment{This could be a figure for your sec 3 (but the bars don't need to be so thick, and you can simply do a stack bar chart so this comparison will be aggregated into a single bar: \url{https://vega.github.io/vega-lite/examples/bar_diverging_stack_transform.html}) For the teaser figure it's more important to show your idea, e.g. show two examples from the same domain, one getting the right doc from DPR and another one benefitting from e.g. BM25, and show how MoR give you right selections on both examples. }

\begin{table}[t]
  \centering
  \small
  \begin{tabular}{lcccc}
    \toprule
    Retriever & NFCorpus & SciDocs & SciFact & SciQ \\
    \midrule
    RepLlama-7b & 38.1 & 18.1 & 76.0 & 65.9 \\ % avg. 48.7
    GritLM-7b & 38.8 & 27.7 & 79.8 & 79.7 \\ % avg. 56.5
    % \midrule
    Route Oracle & 45.1 & 35.1 & 84.4 & 96.6 \\ % avg. 65.3
    % - fine-grained & 49.4 & 40.1 & 89.4 & 98.8 \\ % avg. 69.4 
    \bottomrule
  \end{tabular}
  \caption{{NDCG@20 comparison across different retrievers and datasets. With \texttt{Route Oracle} as an analytical way for mixture, we can get better performance than large language model-based retrievers.}}
  \label{tab:prelim_2}
  % \vspace{-0.2in}
\end{table}

With the experiment suite in hand, we first validate our assumption of performance gain from the comparative advantages among retrievers with diverse queries in the same domain. %in Section~\ref{sec:introduction}. 
We take SciFact~\cite{wadden-etal-2020-fact} as an example to compare the win rate of retriever pairs across all queries. As shown in Figure~\ref{fig:prelim_1}, despite the large gap in overall performance, DPR can still win on 11.7\% queries compared to more advanced TAS-B~\cite{hofstatter2021efficiently}, suggesting a corresponding performance boost if we are able to select the combine the retrievers. Similarly, for retrievers with drastically different characteristics, e.g., supervised DPR and unsupervised SimCSE~\citep{gao-etal-2021-simcse}, despite similar overall performance, there are significant comparative advantages.

We further verify the \textit{folk wisdom} by considering an analytical case of a mixture of retrievers with candidates $L_\rModel$, i.e., query routing. For each query, we consider a simplified case by selecting the best-performing retriever, assuming that the mixture weights just rule out useless retrievers. We denote these analytical results as \texttt{Route Oracle}. 
% In addition, we note that each retriever can potentially be the best at identifying the query-document relations at different scales, e.g., BM25 works well on finding the terminology matching, and SimCSE can work well on capturing sentences with similar semantics. In light of this observation,
% similar to~\citet{cai-etal-2024-mixgr}, for each retriever, we consider its variants with different granularities. We further denote the version as \texttt{Oracle (fine-grained)} (Details in Section~\ref{sec:deep_fusion}). 
Comparing \texttt{Route Oracle} with state-of-the-art LLM retrievers with 7B parameters, as shown in Table~\ref{tab:prelim_2}, we see that the former achieves consistently better performance than GritLM (overall +13.5\%), which suggests the potential gain from the mixture.

% With the awareness of multiple granularities, \texttt{Oracle (fine-grained)} further improves performance, which validates the importance of finer granularities (overall +19.8\% compare to GritLM). 

\section{Mixture of Retrievers.} 
\label{sec:method}

% \swcomment{Before we talk about the mixture of retriever, should there be a section just on the analysis of retriever performances and highlight the need of mixture? Basically sec 4.1 (which also provides an overview of existing retrievers and how they have different advantages and disadvantages) and 4.2? Jumping directly from intro to here could feel a bit abrupt. And in general I think task formation is more useful when it talks about why this task is necessary, rather than just provide definitions of notions. }

Section~\ref{sec:oracle_retrievers} suggests that mixing small retrievers has the potential to outperform large ones. However, \texttt{Route Oracle} operates under a simplified setting. In practice, identifying the ideal retriever to route to is nontrivial, and naive routing overlooks the potential gains from leveraging signals across multiple retrievers.
To address this, we propose a more practical framework: \texttt{MoR} (Mixture of Retrievers), which incorporates a diverse set of signals capturing interactions between queries, retrievers, and documents—without relying on groundtruths.

%on how leveraging a set of BERT-sized retrievers (0.836 B parameters, cumulatively) can potentially achieve comparable performance with large retrievers with 7 billion parameters, in this section, we further seek to build an empirical mixture of retrievers (\texttt{MoR}) with a diverse set of signals modeling the queries, retrievers, and documents. 

% TODO: refactor after deciding either one of the two versions of desiderata
\paragraph{The desiderata.} 
Recall that a retriever $\rModel_i \in L_\rModel$ assigns a relevance score $s_i(\query, \doc_j)$\footnote{The scores are normalized to [0,1] for each retriever.} to each document $\doc_j$, given a query $\query$ and corpus $\corpus= \{\doc_j\}_{j=1}^{|\corpus|}$. This results in a full corpus-sized vector $\retrievalScores_i \in \slReals^{|\corpus|}$.
Each query is then sent to all retrievers, resulting in $N$ relevance scores per document $\doc_j$, where $N = |L_\rModel|$ is the number of retrievers.

Concurrently, we compute a scalar weight for each retriever $\rModel_i$, using a weight allocation function $\MoRModel(\query,\rModel_i,\corpus)$, which estimates the \textit{effectiveness} of $\rModel_i$ for the given query. 
These weights are then used to compute an adjusted relevance score $\tilde{s}(\query, \doc_j)$ for each document $\doc_j$ via a weighted sum over all retrievers:
\begin{align*}
    \tilde{s}(\query, \doc_j) = \sum_{i=1}^N \MoRModel(\query,\rModel_i,\corpus) s_i(\query,\doc_j) ,
\end{align*}
followed by re-ranking of the documents $\{\doc_j\}$ based on the adjusted scores.

A key advantage of this approach is that, due to the zero-shot nature of our geometry-based scores in the embedding space, it naturally supports a \emph{plug-and-play} integration of diverse retrievers, including human information sources that provide ranked documents, as long as appropriate sentence embedding models are available. We explore the collaborative setup further in Section~\ref{sec:human_retriever}.

% We further discuss this efficiency improvement, along with the selection strategy for $L_\rModel$, in Section~\ref{sec:efficiency}.

\subsection{Comprehensive retriever coverage}
\label{sec:deep_fusion}
Inspired by \citet{chen2023dense} and \citet{cai-etal-2024-mixgr}, retrieval indices with different granularities may provide different performance to retrievers, e.g., sub-question index may be paired better with keyword matching, while sentence-level index may work better with semantic embedding-based retrievers. Besides the list of retrievers we introduce in Section~\ref{tab:retriever_comparison}, for each retriever $\rModel_i\in L_\rModel$, we also consider its variants with various kinds of granularity matching. We denote this retriever expansion as a \textbf{deep fusion} of retrievers. 

Specifically, we consider four variants of retrievers with different indices: $R^\text{q-d}$ (original questions and documents), $R^\text{q-p}$ (questions and propositions), $R^\text{sq-p}$ (sub-questions and passage), and $R^\text{sq-p}$ (sub-questions and propositions), following the original notations. To acquire aligned semantic units, we utilize \textit{propositioner} released by \citet{chen2023dense} to break down both queries and documents into atomic units, i.e., sub-questions and propositions (atomic short sentences), respectively.\footnote{For a sentence, \textit{Alice and Bob had coffee together}, the propositions can be \textit{Alice had coffee} and \textit{Bob had coffee}. More details and efficient proposition representation extraction methods can be found in~\cite{chen2023subsentence}.} This means that we extend the number of retrievers by four times ($4N$), with no adaptation on the retrievers needed. We include details and examples of the proposition decomposition in Appendix~\ref{sec:proposition_details}. 

\subsection{Weighing retrievers' effectiveness} 
We explore two complementary weight allocation methods to form $\MoRModel(\query, \rModel_i, \corpus)$: 
\begin{inlinelist}
    \item \textit{Pre-retrieval}: we estimate which retrievers to focus more on by comparing the query embedding to the document embedding centroids generated by each retriever; and 
    \item \textit{Post-retrieval}: we estimate the effectiveness of each retriever by comparing its ranked retrieved documents to the overall corpus distribution.
\end{inlinelist}
% (1) To estimate which retrievers can be down-weighted pre-retrieval, we compare a query to the document embedding centroids created by different retriever; (2) To estimate which retrievers are more promising post-retrieval, we can further compare the ranked retrieved documents with the whole corpus.

\paragraph{Pre-retrieval signals.}
% \fcc{The addition of full stops should be unified in paragraph headers.}
We define pre-retrieval signals as measures of how likely the retriever can perform well on a query before seeking the top relevant documents from the whole corpus. This is closely related to the \textit{when-to-retrieve} problem, where previous work studies from the angles of query token occurrences~\cite {mallen-etal-2023-trust} and model familiarity~\cite [Thrust,][]{zhao2023thrust}.
We extend previous cluster-based familiarity analysis among queries with LLM generators (i.e., readers) to the query-document relations with retrievers. 
% To capture the perplexity of sentence representation models, motivated by CALM~\cite{schuster2022confident}, we consider the layer-wise variance of the transformer. Following our notation in this section,  the signal can be written as: $V_q(q,R)$ = (our implementation of the layer-wise variance).
% \tekc{We say we follow this research but without knowing it and reading it, readers have no idea what it is and what query-query similarity is.}\xrc{adjusted.}
% We leverage the isomorphism of the query-document space to improve Thrust and relieve its cold-start problem. 
% \tekc{So pre-retrieval signal is not only query-specific but also retriever-specific right?}\xrc{yes, it is per-query per-retriever}
Specifically, for the embedding space of each retriever ${\rModel_i}$, we use KMeans Clustering to build K clusters $\{C_1,...,C_k\}$ over the corpus $\corpus$.\footnote{We choose $K$ to be $\max(\text{ceil}(\sqrt[4]{|\corpus|}), 3)$.} 

We design the new pre-retrieval signal as the sum of the vectorized distance from the query vector ($\vec{\query}$) to the cluster centroid vectors ($\{\vec{m}_1,...,\vec{m}_k\}$) (weighed by their sizes).
Then, when the vectorized distance is large, it means that the query vector is distant from all clusters or with similar distance to distant clusters. In the former case, the query is likely an outlier for ${\rModel_i}$, whereas the latter suggests that the retriever can not categorize the query into any type of documents. As a result, we down-weight ${\rModel_i}$. On the other hand, a small distance suggests that the query clearly belongs to one type of document, where we up-weight ${\rModel_i}$ with the reciprocal of the distance.
If we denote the retriever encoded query as $\vec \query$, the pre-retrieval signal is defined as follows:
%$\texttt{Emb}_{\rModel_i}(\query)$
% \begin{align}
%     V_\text{pre}(\query,\rModel_i,\corpus)
%     \triangleq 
%     \bigg\|
%         \sum_{k=1}^K \frac{w_k}{K} \cdot \frac{\vec m_k - \vec \query}{\|\vec m_k - \vec \query\|_2}
%     \bigg\| ,
% \end{align}
\begin{align*}
    V_\text{pre}(\query,\rModel_i,\corpus)
    \triangleq 
    \bigg\|
        \sum_{k=1}^K \frac{|\mathcal{C}_k|}{K} \cdot \hat{v}_k \cdot \frac{1}{\|\vec{v}_k \|_2^2}
    \bigg\| ,
\end{align*}
where $\vec{v}_k=\vec m_k - \vec \query$, the vectorized distance pointing from the query vector to
the centroid embedding $m_k$ of the $k$'th cluster (with a corresponding unit vector $\hat{v}_k$ indicating the direction), and $|\mathcal{C}_k|$ being the size of the $k$'th cluster.
% \swcomment{Worth of simplifying the equations. e.g. it could just be Vpre(Ri | q). In general none of these equations are quite intuitive if we just read the text, but the  euqation itself makes it feel like we are doing something very complex and not clean lol Ideally we would simplify it into something like this: \url{https://chatgpt.com/share/682950f9-3f54-800a-9f68-16e1379ff64f} You can probably ask chatgpt what you could simplify...}
This effectively captures the semantic proximity between the query and the corpus accessible to the retriever and signals how closely the query embedding aligns with the dominant regions of the corpus under $\rModel_i$, with higher values suggesting greater familiarity.

% If we denote the retriever encoding process as $\rModel(\query)$, the pre-retrieval signal is defined as follows:
% \begin{align} 
%         V_\text{pre}(\query,\rModel,\corpus) 
%             \triangleq 
%                     \bigg\|
%                         \frac{1}{K} 
%                         \sum_{k=1}^K 
%                         \frac{| \mathcal{C}_{kl}| }{\|d_{kl}(\query)\|^2}
%                         \cdot
%                         \frac{d_{kl}(\query)}{\|d_{kl}(\query)\|}
%                     \bigg\|,
%         \label{eq:v_pre}
%     \end{align}
% where $d_{kl}(q) \triangleq m_{kl} - \rModel(\query)$ is a vector pointing from $f(q)$ towards the centroid $m_{kl}$, $K$ is the number of clusters\footnote{Following the original work, we choose $K$ to be $\max(\text{ceil}(\sqrt[4]{|\corpus|}), 3)$, where $|\corpus|$ is the size of the corpus.}, $|\mathcal{C}_{kl}|$ denotes the cardinality of the set $\mathcal{C}_{kl}$, and $\|\cdot\|$ denotes $\ell_2$-norm of a vector.

% \input{04-exp}
% \input{05-results}

\begin{table*}[t]
  \centering
  \renewcommand{\arraystretch}{0.95}

  \resizebox{0.9\textwidth}{!}{

  \begin{tabular}{r|*{11}{c}}
    \toprule
    & \multicolumn{2}{c}{NFCorpus} & \multicolumn{2}{c}{SciDocs} & \multicolumn{2}{c}{SciFact} & \multicolumn{2}{c}{SciQ} & \multicolumn{2}{c}{\textbf{Average across datasets}} \\
    \cmidrule(lr){2-3} \cmidrule(lr){4-5} \cmidrule(lr){6-7} \cmidrule(lr){8-9} \cmidrule(lr){10-11}
    & ND@5 & ND@20 & ND@5 & ND@20 & ND@5 & ND@20 & ND@5 & ND@20 & ND@5 & ND@20 \\
    \midrule
    \multicolumn{11}{c}{Unsupervised Retrievers} \\
    \midrule
    BM25 & 37.8 & 30.7 & 14.8 & 19.9 & 64.7 & 69.2 & 91.9 & 92.2 & 52.3 & 53.0 \\
    SimCSE & 16.2 & 13.3 & 7.6 & 9.7 & 27.1 & 31.2 & 62.3 & 67.3 & 28.3 & 30.4 \\
    Contriever & 42.2 & 34.9 & 13.5 & 18.5 & 64.5 & 68.5 & 67.2 & 70.0 & 46.9 & 48.0 \\
    \midrule
    \multicolumn{11}{c}{Supervised Dense Retrievers} \\
    \midrule
    DPR & 25.1 & 20.7 & 7.3 & 10.4 & 31.8 & 37.7 & 60.6 & 64.1 & 31.2 & 33.2 \\
    ANCE & 19.9 & 24.4 & 9.3 & 13.1 & 41.5 & 45.3 & 66.4 & 69.1 & 34.3 & 38.0 \\
    TAS-B & 42.3 & 34.1 & 13.8 & 19.3 & 60.1 & 65.6 & 84.8 & 86.3 & 50.3 & 51.3 \\
    MPNet & 45.6 & 38.7 & 21.3 & 30.3 & 64.9 & 69.4 & 68.0 & 71.9 & 50.0 & 52.3 \\
    GTR & 42.1 & 34.1 & 13.6 & 18.9 & 58.3 & 62.2 & 83.3 & 84.4 & 49.3 & 49.9 \\
    \midrule
    \multicolumn{11}{c}{\textbf{Mixture of Retrievers}} \\
    \midrule
    \texttt{MoR-pre} & 47.7 & 40.4 & 20.9 & 27.5 & 68.7 & 72.8 & 91.4 & 91.6  & 57.2 & 58.1 \\ 
    \texttt{MoR-post} & \textbf{48.0} & \textbf{40.5} & \textbf{21.5} & \textbf{28.1} & 68.9 & 73.2 & \textbf{92.7} & \textbf{92.8} & \textbf{57.8} & \textbf{58.7} \\
    \midrule
    \multicolumn{11}{c}{LLM-based Retrievers (7b parameters)} \\
    \midrule
    RepLLaMA  & 39.8 & 36.4 & 11.9 & 18.2 & 72.5 & 74.1 & 63.3 & 65.9 & 46.9 & 48.7 \\
    GritLM & 47.7 & 38.8 & 20.3 & 27.7 & \textbf{76.9} & \textbf{79.8} & 78.4 & 79.7 & 55.3 & 56.5 \\

    \bottomrule
  \end{tabular}
  }
  \caption{{Performance comparison of different retrievers across datasets using NDCG@5 and NDCG@20 metrics. \textit{Avg.} denotes the macro-average across the tasks. \textbf{Bold} indicates the best performing rows. \texttt{MoR} variants achieve better overall performance than their component retrievers, as well as non-component LLM-based retrievers.}}
  \label{tab:main_performance}
  \vspace{-0.1in}
\end{table*}

\paragraph{Post-retrieval signals.}  
We define post-retrieval signals as measures of how likely the retrieved results are correct, building on a rich line of work in the query performance prediction literature~\cite{10.1145/1277741.1277841,10.5555/2600139,10.1145/3077136.3080665,10.1145/3539618.3592082,10.1145/3673791.3698438}. 
To preserve the zero-shot nature of our approach, we use the Moran coefficient~\cite{10.1145/1277741.1277841}, denoted as $I_{\text{Moran}}(\query,\rModel_i,\corpus)$. This coefficient produces a scalar value that quantifies the correlation among the retrieved documents and has been shown to correlate with retrieval performance. It builds on the cluster-hypothesis \cite{Jardine71cluster}, which posits that closely related documents tend to be relevant to the same query. Therefore, a higher coefficient indicates a greater likelihood that the retrieved documents are relevant to the query.

In addition to measuring relevance among retrieved documents using the Moran coefficient, we also assess their relevance with respect to the entire corpus. Similar to our pre-retrieval signals, we further extend the sum of the vectorized distance from query-document comparison to direct document-document comparison. Suppose that for a query $\query$, the top-ranked retrieved documents are $\corpus_\query$,
\begin{align*} 
        V_\text{post}(\query,\rModel_i,\corpus) 
            \triangleq 
                    \bigg\|
                        \frac{1}{|\corpus_\query|} 
                        \sum_{n=1}^{|\corpus_\query|}
                        V_\text{pre}(\doc_n,\rModel_i,\corpus)
                    \bigg\|,
        \label{eq:v_post}
    \end{align*}

where empirically, we set $|\corpus_\query|=20$. %\xrc{what is the best number; have we tried parametric combination?}

%We compare the individual components of our method as well as other baselines from performance prediction literature~\cite{10.1145/3624918.3625330} in Appendix~\ref{sec:extended_ablation}.% e.g., Reciprocal Rank Fusion (RRF).

\subsection{Parametric Combination.} 
Upon acquiring the individual signals, we design two types of per-query per-retriever weight allocation methods. For pre-retrieval \texttt{MoR-pre}, $\MoRModel_\text{pre}(\query,\rModel_i,\corpus)= V_\text{pre}(\query,\rModel_i,\corpus)$. For post-retrieval \texttt{MoR-post}, we consider the parametric combinations of the signals, i.e., $\MoRModel_\text{post}(\query,\rModel_i,\corpus) = a \cdot V_\text{pre} + b \cdot I_\text{Moran} + c \cdot V_\text{post}$, where coefficients $(a,b,c)$ are hyperparameters. 
Empirically, we select the final set of $(a,b,c)$ for \texttt{MoR-post} as $(0.1, 0.3, 0.6)$.
We use the same set of $a, b, c$ for all queries. We also note that the optimal sets can vary across queries, where query-specific coefficients can constitute an interesting future investigation. More details are in our Limitations (Section \ref{sec:limitations}).
%\swcomment{why these two methods?} \swcomment{and moran looks out of place here}

\section{Experiments and Analyses}

To assess the \texttt{MoR} usefulness, we first compare the retrieval and RAG performance of \texttt{MoR} with various supervised and unsupervised retrievers in Section~\ref{sec:main_results}, using retrievers in Section~\ref{sec:experiment_settings}.\footnote{RepLLaMA and GritLM are not considered as components of $L_\rModel$ for \texttt{MoR} due to efficiency reasons.} Then, we experiment on how the mixture framework can help incorporate specialized non-oracle human information sources (Section~\ref{sec:human_retriever}). We further study \texttt{MoR} efficiency in Section~\ref{sec:efficiency}.
Details of the list retrievers included in \texttt{MoR} are in Section~\ref{sec:experiment_settings}. 

\subsection{Main results: \texttt{MoR} is Effective}
\label{sec:main_results}
% \swcomment{restructure the result section so each paragraph starts with a bold conclusion / overview before adding more elaboration}
\paragraph{\texttt{MoR} improves retrieval performance.} 
% From Table~\ref{tab:main_performance}, we can observe that, similar to the findings of \cite{thakur2021beir}, BM25 remains a strong baseline in scientific retrieval tasks.\swcomment{is this sentence on bm25 necessary?}
From Table~\ref{tab:main_performance}, comparing \texttt{MoR} with their component retrievers, we can observe that our zero-shot signals achieve good performance on combining the query-document scores from different retrievers. Across different tasks, \texttt{MoR-pre} achieves improved performance over various unsupervised and supervised retrievers, which demonstrates the significance of our proposed $V_\text{pre}$ without searching the most similar documents from the corpus. Through considering $I_\text{Moran}$ and $V_\text{post}$, \texttt{MoR-post} achieves even better performance than the pre-retrieval variant, with a relative 10.8\% and 12.2\% performance improvement on NDCG@20 over the best unsupervised and supervised components, respectively.

With further comparison to LLM-based retrievers, we can observe that a mixture of smaller retrievers can surpass the performance of large ones, with an overall +3.9\% better relative NDCG@20 improvement over GritLM. Across different tasks, \texttt{MoR-post} achieves better performance than GritLM on NFCorpus, SciFact, and SciQ with 5 or 20 chunks considered. For SciFact, \texttt{MoR-post} achieves better performance than its components and comparable performance to GritLM.

We show detailed qualitative examples in Appendix~\ref{sec:qualitative_analysis_appendix} in the appendix to reveal further details on how \texttt{MoR} works, e.g., selecting the correct one when most retrievers fail or ignoring the wrong output from the overall best-performing retriever.

\begin{table}[t]
  \centering
  \small
  \begin{tabular}{r | c c | c c}
    \toprule
    & \multicolumn{2}{c}{SciFact} & \multicolumn{2}{c}{SciQ} \\
    %\cmidrule(lr){2-3} \cmidrule(lr){4-5} \cmidrule(lr){6-7} \cmidrule(lr){8-9} \cmidrule(lr){10-11}
    & EM@1 & EM@3 & EM@1 & EM@3 \\
    \midrule
    BM25 & 67.4 & 71.5 & 64.1 & 68.8 \\
    MPNet & 66.9 & 75.6 & 57.0 & 61.3 \\
    RepLLaMA & 45.9 & 65.7 & 56.2 & 64.1 \\
    GritLM & 66.9 & 77.3 & 61.7 & 66.8 \\
    \midrule
    \texttt{MoR-pre} & 68.0 & 74.4 & 62.9 & 67.2 \\
    \texttt{MoR-post} & \textbf{72.9} & \textbf{77.9} & \textbf{66.8} & \textbf{68.8} \\
    \bottomrule
  \end{tabular}
  \caption{\small{Retrieval augmented generation performance on SciFact and SciQ with top-1 (EM@1) or 3 (EM@3) chunks fed into the reader model. EM denotes using Exact Match as the metric. \textbf{Bold} indicates the best performing row.}}
  \label{tab:rag_performance}
  % \vspace{-0.2in}
\end{table}

\paragraph{\texttt{MoR} improves RAG performance.} Besides the gain on retrieval performance. We further validate the impact of \texttt{MoR} on RAG with SciFact and SciQ that have downstream generation tasks, i.e., fact-checking and question answering, respectively.
To do so, we use the standard RAG pipeline to feed the doc chunks retrieved into a reader model Llama-3-8B-Instruct~\cite{DBLP:journals/corr/abs-2307-09288}. We use Exact Match (EM@K) to measure the retrieval-augmented generation performance, where K denotes the number of retrieved chunks considered.

From Table~\ref{tab:rag_performance}, we can observe that, similar to the retrieval performance, \texttt{MoR} shows consistent performance improvement over baselines, which demonstrates the effectiveness of \texttt{MoR} on RAG even without utilizing any signals from the final generation beforehand to assign weights. Yet, post-presentation signals from downstream generation, e.g., accuracy, can still be important future work beyond our current scope.

\subsection{Human as a Retriever } 
\label{sec:human_retriever}

\begin{table}[t]
  \centering
  \small
  \setlength{\tabcolsep}{3pt}
  \begin{tabular}{r|cccc}
    \toprule
    Domain Weights & Medicine  & Psychology & CS & Eng. \\
    \midrule
    Medicine Expert & \textbf{0.6} & 0.0  & 0.0 & 0.0 \\
    Psychology Expert & 0.0 & \textbf{0.8}  & 0.0 & 0.0 \\
    CS Expert & 0.0 & 0.1  & \textbf{0.8} & 0.1 \\
    Engineering Expert & 0.2  & 0.0  & 0.1 & \textbf{0.7} \\
    \bottomrule
  \end{tabular}
  \caption{{Averaged weights of $V_\text{post}$ assigned to each simulated human expert on queries from different domains. Simulated experts can conduct Oracle retrieval on their corresponding domains. \textbf{Bold} denotes the highest weights for each query domain (column).}}
  \label{tab:human_retriever_weights}
  % \vspace{-0.1in}
\end{table}

\begin{table}[t]
  \centering
  \small
  \setlength{\tabcolsep}{3pt}

  \begin{tabular}{rcccc}
    \toprule
    NDCG@20 & Medicine  & Psychology & CS & Eng. \\
    \midrule
    GritLM & 42.3  & 48.1  & 20.5 & 42.2 \\
    \texttt{MoR-post} & 44.3 & 44.4  & 20.7 & 43.2 \\ %38.2
    \midrule
    Human Experts & 71.2 & 53.5 & 40.0 & 66.2 \\ % 57.7
    \texttt{MoR}+Humans & 87.2  & 91.5  & 94.1 & 94.3 \\ %avg: 91.7
    \bottomrule
  \end{tabular}
  \caption{{Performance of each domain with GritLM, MoR, and MoR+Humans. Each human expert can conduct Oracle retrieval on the corresponding domain, but random retrieval for others. \textit{Humans Experts} denotes the retrieval performance when we assign equal weights to each one's ranks.}}
  \label{tab:human_retriever_performance}
  % \vspace{-0.2in}
\end{table}

% \swcomment{Move this before ablation? And similar to intro, maybe make a stronger connection argument -- Saying we are doing stress test to see if this allow us to generalize to all kinds of retrievers, including also using humans as retrievers}

In our main experiments, we mainly consider the mixture of sparse and dense retrievers, such as BM25 and DPR. However, for complex scientific domain retrieval, human experts shall also be considered as an important source of information. 
In this section, we conduct a stress test of \texttt{MoR} to explore the potential to generalize to all kinds of information sources, including using humans as retrievers.
Specifically, we simulate human experts using four domains from the original SciDocs splits\footnote{\url{https://github.com/allenai/scidocs}. General engineering denotes non-CS engineering topics.}, including Medicine, Psychology, Computer Science (CS), and general Engineering (Eng.). We construct 4 corresponding human experts, where each expert can conduct Oracle retrieval on its own domain (gold documents are ranked top, and the rest and ranked with their relevance to the gold documents). For queries from other domains, the simulated experts will output random ranks. We use MPNet to encode the ranked documents into their semantic representation. % of the documents with biases to their own domains.

\paragraph{\texttt{MoR} assigns reasonable weights to simulated human retrievers.} With the setting above, we study whether our $V_\text{post}$ can delegate reasonable weights to the human experts. From Table~\ref{tab:human_retriever_weights}, we can observe that experts are consistently weighted high in their corresponding domains (which means their ranks will be counted more in the final \texttt{MoR} ranks) and low in non-expert domains. The reasonable weights highlight the effectiveness of $V_\text{post}$ in a controlled setting, and show that our method has the potential to help estimate human trustworthiness from a given corpus.

\paragraph{\texttt{MoR} achieves improved performance to simulated human retrievers.}
We see from Table~\ref{tab:human_retriever_performance} that without simulated human experts, \texttt{MoR} achieves on par performance with GritLM on SciDocs retrieval task for these specific domains. However, through delegating reasonable weights to the human experts and include their ranks into consideration, \texttt{MoR} achieves consistent best performance across all domains, outperforms aggregating the human experts (+58.9\% relatively), which shows its potential of including non-oracle humans as information sources in collaborative tasks.

\subsection{Efficiency} % and Performance Trade-offs
\label{sec:efficiency}

\begin{table}[t]
  \centering
  % \small
  \scriptsize
  \resizebox{\linewidth}{!}{
  \begin{tabular}{l|lc}
    \toprule
    Best of & Retrievers & NDCG@20\\
    \midrule
    2 & Contriever, GTR & 92.6 \\
    3 & Contriever, GTR. MPNet & 92.8 \\
    4 & SimCSE, DPR, Contriever, GTR & 92.9 \\
    % 5 & SimCSE, DPR, Contriever, GTR, MPNet & 93.0 \\
    \bottomrule
  \end{tabular}
  }
  \caption{
  Results mixing a subset of \emph{X} retrievers for SciQ, with \texttt{MoR-post}. Mixing the best 2 achieves comparable results to our original mixture of 8 in Table~\ref{tab:main_performance}.
  }
  \label{tab:best_retriever_suite}
  % \vspace{-0.1in}
% Best retriever combination when selecting X retrievers from $L_\rModel$ (Best of X) for \texttt{MoR}. ND@20 denotes the NDCG@20 performance of SciQ with \texttt{MoR-post}.}{
\end{table}

\begin{figure}[t]
    %\vspace{-0.2in}
    \centering
    \includegraphics[clip,trim={0.2cm 0.2cm 0.2cm 0.2cm},width=0.9\linewidth]{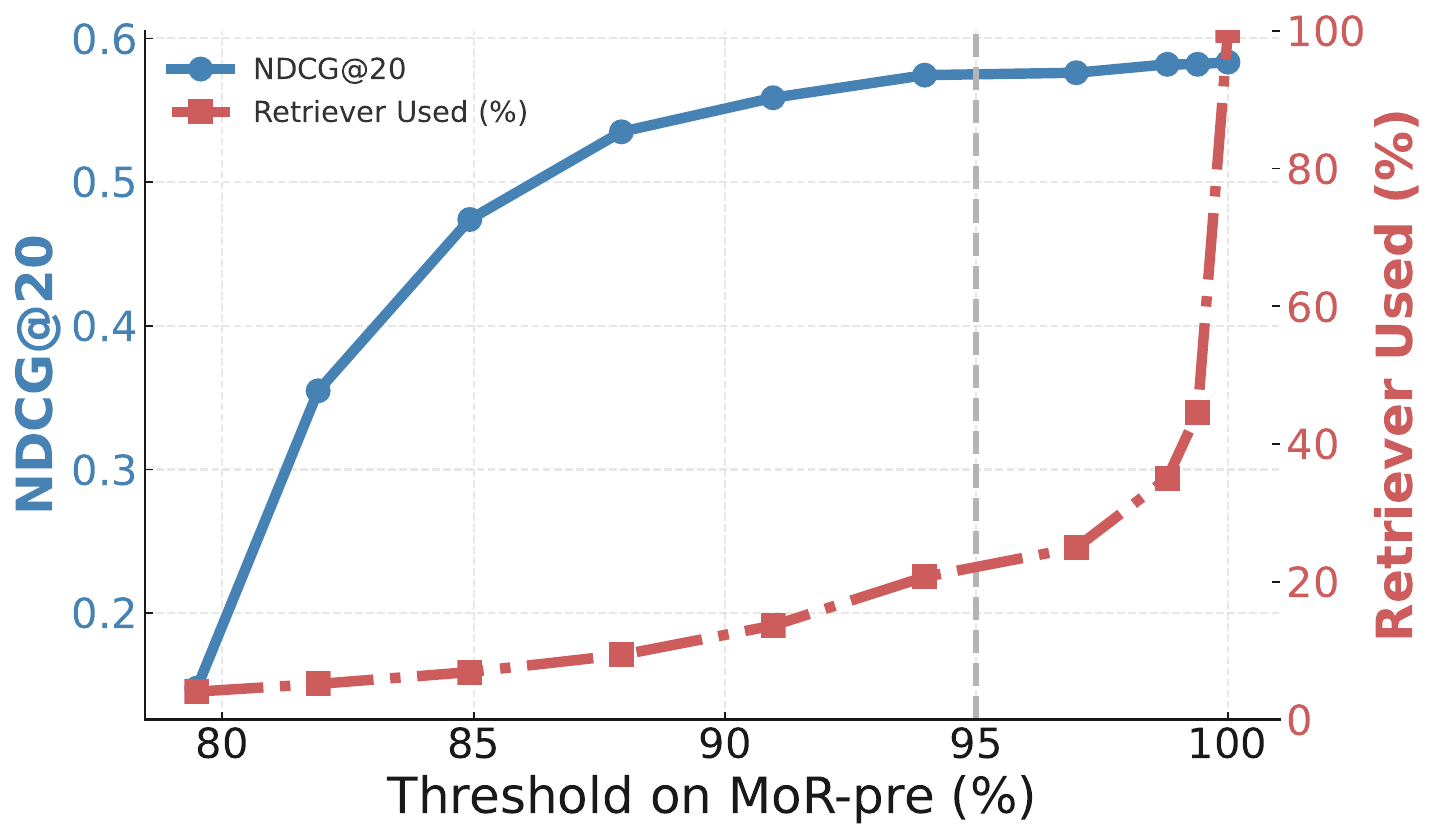}
    % \vspace{-0.1in}
\caption{Performance and equivalent retriever used (\%) at different thresholds on \texttt{MoR-pre} weights. Threshold (\%) at 95 percentile between min and max weights allows \texttt{MoR} to maintain most performance with only around 20\% equivalent retriever use.}
\label{fig:threshold}
\vspace{-0.1in}
\end{figure}

In addition to evaluating performance, we also consider efficiency as a key metric for assessing \texttt{MoR}. Following standard RAG conventions, corpus embeddings are pre-computed during offline preparation. As a result, the primary computational cost at test time arises from encoding the input queries and searching for the most relevant corpus entries --- Both increases linearly with the number of dense retrievers we include. 
To improve the \texttt{MoR}'s efficiency, we pose the question: \textit{Which retrievers should we include?} 
{
As a stress test, we exhaustively enumerated all possible subsets of $X$ retrievers drawn from $L_\rModel$, and examine the performance of the best combinations.
As shown in Table~\ref{tab:best_retriever_suite}, mixing just two retrievers can already yield performance comparable to mixing all eight.
Notably, the best-performing pair is \emph{not} composed of the individually top-performing retrievers (MPNet and BM25, per Table~\ref{tab:main_performance}).
This suggests that subset selection can significantly improve \texttt{MoR} efficiency without sacrificing much performance; However, effective selection should prioritize complementarity over absolute performance.

Building on this, we explore a natural extension to \texttt{MoR}: pre-rejecting certain retrievers using \texttt{MoR-pre} weights, calculated \emph{before} observing actual query-document interactions.
Concretely, we normalize \texttt{MoR-pre} scores into percentiles and apply a rejection threshold, discarding low-ranking retrievers accordingly.
Figure~\ref{fig:threshold} shows this early rejection strategy is effective: at a 95th percentile threshold, \texttt{MoR} maintains strong performance while using, on average, around only 20\% of retrievers per query.
}

\subsection{Ablation Study}
\label{sec:ablation}

\begin{table}[t]
  \centering
  \small
  \begin{tabular}{l|cc|c}
    \toprule
     & Granularity & Retriever & \multicolumn{1}{c}{Avg.} \\
    & Merge & Merge & ND@20  \\
    \midrule
    % RRF & max & mean & 54.2\\
   \texttt{MoR}& max & mean & 46.3\\
   \texttt{MoR} & mean & mean & 35.2\\
   \texttt{MoR}& mean & $V_\text{pre}$ & 50.7  \\
   \texttt{MoR}& none & $V_\text{pre},I_\text{Moran},V_\text{post}$ & 56.0 \\
    \midrule
    \texttt{MoR-pre} & \multicolumn{2}{c|}{$V_\text{pre}$} & 58.1  \\
    \texttt{MoR-post} & \multicolumn{2}{c|}{$V_\text{pre},I_\text{Moran} ,V_\text{post}$} &  58.7  \\
    \bottomrule
  \end{tabular}
  \caption{{Ablation studies with retrieval performance of Avg. NDCG@20 metrics. Granularity Merge denotes the way to merge the scores of the same retriever operated in different granularities. Retriever Merge denotes the way to merge scores from different retrievers.}}
  \label{tab:ablation_study}
  % \vspace{-0.2in}
\end{table}

In Table~\ref{tab:ablation_study}, we further validates the design choices of \texttt{MoR} with ablations. Specifically, we consider variants of \texttt{MoR} with different treatments on the weight delegation on various retrievers (Retriever Merge), as well as their variants on different granularities (Granularity Merge). From the comparison of the average performance across our task suite, Table~\ref{tab:ablation_study} presents that the signals we fetched show significantly better performance than adhoc mean or max over the retriever scores for each query. Besides, the no granularity merge variant shows the effectiveness of including the deep fusion.
Further comparison of individual components of our method as well as other alternatives, e.g., Reciprocal Rank Fusion (RRF), can be found in Appendix~\ref{sec:extended_ablation}.

\section{Conclusion}
In this paper, we propose to construct a mixture of retrievers (\texttt{MoR}) to improve the retrieval generalizability on diverse and complex retrieval tasks, leveraging the comparative advantages among small-scaled retrievers. To do so, we propose to use deep fusion considering multi-granularities, as well as design various pre-/post-retrieval signals to weight the outputs of each retriever for each query. Experiments on various tasks and settings validate the consistent and robust performance improvement of \texttt{MoR} compared to its component retrievers and SOTA LLM-based retrievers. Extensive analysis further sheds light on how \texttt{MoR} can potentially incorporate human information sources and be implemented with improved efficiency. We will open-source our code at \url{https://github.com/Josh1108/MixtureRetrievers}.

% \clearpage

\section*{Limitations}
\label{sec:limitations}
\paragraph{Post-presentation Signals.}
In our main experiments, we design various effective signals before and after conducting retrieval. However, there is one source of signals that can be an important future direction - the end-to-end performance after utilizing the retrieved documents, e.g., the exact match performance on question answering or the user satisfaction. We denote such signals as post-presentation signals, which can potentially extend the design from the retrieval perspective to the RAG~\cite{jiang-etal-2023-active} or agentic retrieval~\cite{Search-o1} perspectives.

\paragraph{Supervised Methods.} In this paper, we focus on proposing the \texttt{MoR} architecture and design of unsupervised signal sources to allow potential extensiveness to future retrievers. However, there is a zoom for potential supervised methods at different stages of MoR. First, intuitively, all the per query per retriever weight can be learned through a neural network, considering $q,R,D',D$ as the inputs given a small set of training data, with the pre-computed optimal weights. Besides, our current parametric combination uses one universal set of coefficients for different signals. Another neural network can be used to compute the coefficients per retriever or per query with the query embeddings as the inputs, which can potentially help model the different kinds of complexity among queries, i.e., the complexity can come from comprehending the query itself (captured by layer-variance) or the lack of good entries (captured by post-retrieval signals).

\section*{Acknowledgments}
The authors thank Tong Chen, Yuhao Zhang, Haoyang Wen, Hongming Zhang, Sihao Chen, and Ben Zhou for their insights into design and evaluation choices. The authors also thank the constructive discussions with colleagues from CMU WInE Lab.
Xinran Zhao is supported by the ONR Award N000142312840.
To Eun Kim is supported by NSF grant 2402874.
This work is supported by the OpenAI Research Credit program, the Amazon AI Research Gift Fund, and the Gemma Academic Program GCP Credit Award. Any opinions, findings and conclusions or recommendations expressed in this material are those of the authors and do not necessarily reflect those of the sponsors.

\bibliography{XX-references}

\clearpage

\appendix
\section{Appendix}
\label{sec:appendix}

\subsection{Experiment Details}
\label{appendix:experimental_details}

\paragraph{Statistics of the data.}
Table~\ref{tab:dataset_stats} presents the statistics of four datasets used in our experiments, together with the statistics of the decomposed queries and documents, i.e., sub-queries and propositions.
Specifically, due to the max-length requirement for some dense retrievers such as DPR~\cite{karpukhin-etal-2020-dense}, we split one document in the original dataset into several chunks containing a maximum of 128 words.
In this way, we can avoid the loss of information caused by context overflow.
The retrieved chunk serves as a reference to locate the corresponding document in the original dataset for evaluation.

\begin{table*}[t]
    \centering
    \resizebox{\textwidth}{!}{%
    \begin{tabular}{lcccc}
        \toprule
        \textbf{Statistic (\#)}  & NFCorpus \cite{boteva2016full} & SciDocs \cite{cohan-etal-2020-specter} & SciFact \cite{wadden-etal-2020-fact} & SciQ \cite{welbl-etal-2017-crowdsourcing} \\
        \midrule
        Query & \numprint{1016} & \numprint{1000} & \numprint{1109} & \numprint{884} \\
        Multi-subquery queries & \numprint{641} & \numprint{205} & \numprint{283} & \numprint{252} \\
        Subqueries & \numprint{3337} & \numprint{522} & \numprint{614} & \numprint{874} \\ 
        \midrule
        Documents  & \numprint{3633} & \numprint{25657} & \numprint{5183} & \numprint{12241} \\
        Propositions & \numprint{67110} & \numprint{351802} & \numprint{87190} & \numprint{91635} \\
        \bottomrule
    \end{tabular}
    }%
    \caption{Statistics for the NFCorpus, SciDocs, SciFact, and SciQ datasets. We note that these statistics have been adjusted to prevent proposition/sub-question decomposition errors.}\label{tab:dataset_stats}
\end{table*}

\begin{table*}[t]
\centering
\resizebox{\textwidth}{!}{%
\begin{tabular}{@{}l l r@{}}
\toprule
\textbf{Model} & \textbf{HuggingFace Checkpoint} & \textbf{Params} \\
\midrule
SimCSE~\cite{gao-etal-2021-simcse} & \texttt{princeton-nlp/unsup-simcse-bert-base-uncased} & 110M \\
Contriever~\cite{izacard2022unsupervised} & \texttt{facebook/contriever} & 110M \\
\multirow{2}{*}{DPR~\cite{karpukhin-etal-2020-dense}}
& \texttt{facebook/dpr-ctx\_encoder-multiset-base} & \multirow{2}{*}{110M} \\
& \texttt{facebook/dpr-question\_encoder-multiset-base} & \\
\multirow{2}{*}{ANCE~\cite{xiong2021approximate}}
& \texttt{castorini/ance-dpr-context-multi} & \multirow{2}{*}{110M} \\
& \texttt{castorini/ance-dpr-question-multi} & \\
TAS-B~\cite{hofstatter2021efficiently} & \texttt{sentence-transformers/msmarco-distilbert-base-tas-b} & 66M \\
GTR~\cite{ni-etal-2022-large} & \texttt{sentence-transformers/gtr-t5-base} & 220M \\
% GTE-Qwen2-1.5B-instruct~\cite{li2023towards} & \texttt{Alibaba-NLP/gte-Qwen2-1.5B-instruct} & 1.5B \\
% E5-Mistral-7B-instruct~\cite{wang-etal-2024-improving-text} & \texttt{intfloat/e5-mistral-7b-instruct} & 7B \\
MPNet~\cite{song2020mpnet} & \texttt{sentence-transformers/all-mpnet-base-v2} & 110M \\
RepLLaMA~\cite{ma2023finetuningllamamultistagetext} & \texttt{castorini/repllama-v1-7b-lora-passage} & 7B \\
GritLM~\cite{muennighoff2024generative} & \texttt{GritLM/GritLM-7B} & 7B \\
\bottomrule
\end{tabular}}
\caption{Model checkpoints released on HuggingFace and model parameters. For DPR and ANCE, the parameter count is shared across the dual encoders.}
\label{tab:model_checkpoints}
\end{table*}

\paragraph{Infrastructure.}
We conduct experiments on a Google Cloud Platform instance equipped with 8×NVIDIA L4 GPUs, each with 24 GB of memory.

\paragraph{Setup and Hyperparameters.}
For retrieval augmented generation, we use a temperature of 0.1 and top\_p value of 0.7 across all tasks.

\subsection{Propositionizer details and decompositions}
\label{sec:proposition_details}
\textit{Propositioner}~\cite{chen2023dense}\footnote{\url{https://huggingface.co/chentong00/propositionizer-wiki-flan-t5-large}} is an off-the-shelf model for query and document decomposition.
Using Wikipedia as the dataset, this model distills GPT-4’s~\cite{achiam2023gpt} decomposition capacity into Flan-T5-Large~\cite{chung2024scaling}.

Propositioner breaks down queries and documents into fundamental components—subqueries and propositions, respectively. Each proposition (or subquery) is expected to satisfy the following three key criteria~\cite{min-etal-2023-factscore}:

\begin{itemize}[leftmargin=*, itemsep=0em]
    \item It should express a distinct semantic unit, contributing to the overall meaning when considered with others.
    \item It must be atomic and indivisible.
    \item Following \citet{choi2021decontextualization}, each proposition should be self-contained and contextually complete, incorporating all necessary information, such as resolved coreferences, for unambiguous interpretation.
\end{itemize}

The example of subqueries and propositions is listed in Figure \ref{fig:netosis-decomposition}.

\begin{figure}[H]
\centering
\begin{tcolorbox}[colback=gray!5, colframe=black!30, boxrule=0.3pt, 
    left=1mm, right=1mm, top=0.5mm, bottom=0.5mm, 
    before skip=0pt, after skip=0pt, width=\columnwidth]
\textbf{Query:} Citrullinated proteins externalized in neutrophil extracellular traps act indirectly to perpetuate the inflammatory cycle via induction of autoantibodies.
\vspace{-0.3em}
\begin{itemize}[leftmargin=*, itemsep=0.2em, parsep=0em]
    \item \textbf{Subquery-0}: Citrullinated proteins are externalized in neutrophil extracellular traps.
    \item \textbf{Subquery-1}: Citrullinated proteins act indirectly to perpetuate the inflammatory cycle.
    \item \textbf{Subquery-2}: The inflammatory cycle is perpetuated via induction of autoantibodies.
\end{itemize}

\textbf{Document:} In humans, RNA blot analysis revealed that Golli-MBP transcripts were expressed in fetal thymus, spleen, and human B-cell and macrophage cell lines, as well as in fetal spinal cord. These findings clearly link the expression of exons encoding the autoimmunogen/encephalitogen MBP in the central nervous system to cells and tissues of the immune system through normal expression of the Golli-MBP gene. They also establish that this genetic locus, which includes the MBP gene, is conserved among species, providing further evidence that the MBP transcription unit is an integral part of the Golli transcription unit and suggest that this structural arrangement is important for the genetic function and/or regulation of these genes.
\begin{itemize}[leftmargin=*, itemsep=0.2em, parsep=0em]
\item \textbf{Proposition-0:} The human myelin basic protein (MBP) gene is contained within a 179-kilobase transcription unit.
\item \textbf{Proposition-1:} Golli-MBP transcripts are expressed in fetal thymus, spleen, human B-cell lines, macrophage cell lines, and fetal spinal cord.
\item \textbf{Proposition-2:} Expression of MBP-encoding exons in the central nervous system is linked to immune-system cells and tissues through normal Golli-MBP expression.
\item \textbf{Proposition-3:} The genetic locus that includes the MBP gene is conserved across species.
\item \textbf{Proposition-4:} The MBP transcription unit is an integral part of the larger Golli transcription unit.
\item \textbf{Proposition-5:} The structural arrangement of the MBP and Golli transcription units is important for the genetic function and/or regulation of these genes.
\end{itemize}
\end{tcolorbox}
\caption{Example of query and document decomposition with \textit{Propositioner}.}
\label{fig:netosis-decomposition}
\end{figure}

\paragraph{RAG experiment details.}
The RAG templates used for SciFact and SciQ are listed below.
For SciQ, we convert the multiple-choice questions into open-ended questions.

\begin{tcolorbox}
    Given the knowledge source: \textit{context} \textbackslash\textbackslash n Question: \textit{query} \textbackslash\textbackslash n Reply with one phrase. \textbackslash\textbackslash n Answer:

\end{tcolorbox}

Since SciFact is a fact-checking task, we evaluate whether LLMs can accurately predict the relationship between a claim and a given context.
The template used for SciFact is as follows:
\begin{tcolorbox}
Context: \{\textit{context}\} \textbackslash\textbackslash n Claim: \{\textit{query}\} \textbackslash\textbackslash n For the claim, the context is supportive, contradictory, or not related? \textbackslash\textbackslash n Options: (A) Supportive (B) Contradictory (C) Not related \textbackslash\textbackslash n Answer:")
\end{tcolorbox}

% \subsection{Extended Human as a retriever analysis} 
% \label{sec:extended_human_retriever}
\subsection{Extended Retriever Descriptions}
\label{sec:appendix_retriever_details}

We consider both sparse and various BERT-sized~\cite{devlin-etal-2019-bert} dense retrievers as candidates $L_\rModel$ to create our mixture of retrievers:
\begin{itemize}[leftmargin=*, itemsep=0em, parsep=0pt, topsep=1pt]
    \item BM25 \citep{10.1145/2682862.2682863} is a traditional key-word matching based sparse retriever that shows good zero-shot performance~\cite{thakur2021beir}. We use TF-IDF vectors to conduct the vector space operations for BM25.
    \item SimCSE \citep{gao-etal-2021-simcse} employs a BERT-base \cite{devlin-etal-2019-bert} encoder trained on self-supervised contrastive signals on Wikipedia sentences.
    \item Contriever \citep{izacard2022unsupervised} is an unsupervised retriever evolved from a BERT-base encoder, contrastively trained on segments from unlabelled web and Wikipedia documents.
    \item DPR \citep{karpukhin-etal-2020-dense} is built with a dual-encoder BERT-base architecture, finetuned on a suite of open-domain datasets with labels, such as SQuAD \cite{rajpurkar-etal-2016-squad}.
    \item ANCE \citep{xiong2021approximate} extend DPR with a training scheme of Approximate Nearest Neighbor Negative Contrastive Estimation (ANCE).
    \item TAS-B \citep{hofstatter2021efficiently} is a dual-encoder BERT-base model distilled from ColBERT on MS MARCO \cite{nguyen2016ms}. %a cross-attention model
    \item GTR \citep{ni-etal-2022-large} is a T5-base encoder, focusing on generalization, pre-trained on unlabeled QA pairs, and fine-tuned on labeled data including MS MARCO.
    \item MPNet~\cite{song2020mpnet} is a BERT-alike model with advanced pre-training strategy. We use it as a sentence transformer to serve as a retriever following~\cite{reimers-gurevych-2019-sentence}.
    
\end{itemize}

As depicted in the descriptions, the above retrievers vary in parameter sizes, backbone architectures, and training signals. In addition, we also set a competitive performance reference with various large-language-model-based retrievers with 7 billion parameters. 

\begin{itemize}[leftmargin=*, itemsep=0em, parsep=0pt, topsep=1pt]
    \item RepLLaMA~\citep{ma2023finetuningllamamultistagetext} extends the dual-encoder retriever training pipeline of DPR to LLaMA~\cite{touvron2023llama}, which demonstrates advanced performance and long-context generalizability.
    \item GritLM~\citep{muennighoff2024generative} proposes to jointly train language models with generative and embedding tasks, which leads to great performance on MTEB~\cite{muennighoff2022mteb}.

\end{itemize}

\paragraph{Model Huggingface checkpoints and size.}
Table \ref{tab:model_checkpoints} lists model checkpoints on Huggingface and the model sizes.
Our experiments cover modern retrievers from different architectures (e.g., BERT~\cite{gao-etal-2021-simcse} or LLaMA~\cite{ma2023finetuningllamamultistagetext}) and a wide range of model size, from 66M to 7B.

\subsection{Extended design choice ablation} 
\label{sec:extended_ablation}

\begin{table*}[t]
  \centering
  \small
  \begin{tabular}{l*{11}{c}}
    \toprule
    & \multicolumn{2}{c}{NFCorpus} & \multicolumn{2}{c}{SciDocs} & \multicolumn{2}{c}{SciFact} & \multicolumn{2}{c}{SciQ} & \multicolumn{2}{c}{Avg.} \\
    \cmidrule(lr){2-3} \cmidrule(lr){4-5} \cmidrule(lr){6-7} \cmidrule(lr){8-9} \cmidrule(lr){10-11}
    & ND@5 & ND@20 & ND@5 & ND@20 & ND@5 & ND@20 & ND@5 & ND@20 & ND@5 & ND@20 \\
    \midrule
    \multicolumn{11}{c}{pre-retrieval signals} \\
    \midrule
    Perf. Nor. & 40.1 & 34.2 & 13.9 & 20.2 & 45.3 & 52.4 & 21.0 & 32.8 & 30.1 & 34.9 \\
    Layer Var. & 40.0 & 34.2 & 14.0 & 20.3 & 34.5 & 43.8 & 20.6 & 32.4 & 27.3 & 32.7 \\
    Clustering & 42.1 & 35.2 & 15.5 & 21.4 & 51.0 & 57.3 & 38.7 & 48.6 & 36.8 & 40.6 \\
    Thrust & 42.8 & 35.6 & 15.7 & 21.3 & 56.9 & 61.8 & 70.0 & 74.3 & 46.3 & 48.2 \\
    $V_\text{pre}$ & 47.7 & 40.4 & 20.9 & 27.5 & 68.7 & 72.8 & 91.4 & 91.6 & 57.2 & 58.1 \\
    \midrule
    \multicolumn{11}{c}{post-retrieval signals} \\
    \midrule
    Score Var. & 40.0 & 34.0 & 13.7 & 19.8 & 41.1 & 49.0 & 20.7 & 30.9 & 28.9 & 33.4 \\
    Rep. Var.  & 39.7 & 34.3 & 14.5 & 20.7 & 42.0 & 49.8 & 21.3 & 32.2 & 29.4 & 34.2 \\
    RRF & 44.6 & 37.7 & 17.0 & 24.1 & 64.2 & 69.2 & 84.2 & 85.6 & 52.5 & 54.2 \\
    \midrule
    Moran & 42.6 & 34.8 & 15.0 & 20.5 & 42.9 & 50.7 & 26.4 & 35.3 & 31.7 & 35.3 \\
    $V_\text{post}$ & 47.7 & 40.4 & 20.8 & 27.3 & 68.8 & 72.9 & 91.5 & 91.9 & 57.2 & 58.1 \\
    $V_\text{post}$ no D.F. & 47.3 & 39.7 & 20.3 & 27.1 & 69.3 & 73.2 & 82.2 & 83.9 & 54.8 & 56.0 \\
    \bottomrule
  \end{tabular}
  \caption{Performance comparison of different signals across datasets using NDCG@5 and NDCG@20 metrics. $V_\text{post}$ no D.F. denotes our method without the deep fusion component.}
  \label{tab:ablation_performance}
\end{table*}

% We compare the performance of our methods with various supervised and unsupervised retrievers in Table~\ref{tab:main_performance}, where the details of MoR-pre and MoR-post are described in Section~\ref{sec:design}. We can observe that we achieve equivalent or better performance than the SOTA GritLM. \xrc{put the absolute numbers and compare with component retrievers and large retrievers}. \xrc{eventually, the details would be put into the appendix. Here we just list the details.}

In Section~\ref{sec:main_results}, we show the performance of MoR-pre and MoR-post, which are parametric combinations of various signals extracted from queries and documents. In this section, we take a closer look at the individual components used in MoR, i.e., $V_\text{pre}$, Moran, and $V_\text{post}$, as well as other baseline signals. For $V_\text{post}$, we also compare with the version without deep fusion of multiple granularities (denoted as no D.F.). 
Following our intuition and design in Section~\ref{sec:method}, for pre-retrieval signals, we consider:

\begin{itemize}[leftmargin=*, itemsep=0em]

    \item Performance Normalization (Perf. Norm.): We consider the simplest baseline on a development set that contains 100 queries to compute the retrieval performance of each retriever and weight each of them with the normalized scores (higher performance indicates higher weights).

    \item Layer-wise Variance (Layer Var.): motivated by \cite{schuster2022confident}, we consider the layer-wise variance of the retriever on the first, middle, and last layers at each neuron to indicate how much computation is required for the retriever to process a specific query. The reciprocal of the variance is then considered as the weight of each query per retriever.

    \item Clustering: We consider the variance of the cluster centroids of the vectors of the corpora that are extracted from each retriever. The reciprocal of variance is then used as the weight.

    \item Thrust~\cite{zhao2023thrust}: we utilize the original implementation of the target query and 300 sample queries to compute the Thrust score representing the retriever familiarity of the queries. The scores are then normalized to serve as the weights, where lower familiarity scores denote less weight.

\end{itemize}

For post-retrieval signals, we consider:

\begin{itemize}[leftmargin=*, itemsep=0em]

    \item Score and Representation Variance: following ~\cite{10.1145/3624918.3625330}, we utilize the relations among the top retrieved documents to conduct the performance prediction. For score variance (Score Var.), we consider the variance of the scores of top-x retrieved documents. Similarly, for representation variance (Rep. Var.), we consider the variance of their embeddings. The reciprocal of variance is then used as the weight, where lower variance denotes higher weights.
    % \tekc{clearly indicate lower the better}
    
    \item Reciprocal Rank Fusion (RRF; \citealt{cormack2009reciprocal}): following~\cite{langchain2025}, we consider RRF as a direct baseline to combine the document ranks of the retrievers.

    % \item Moran Index: following ~\cite{10.1145/1277741.1277841}, we compute the Moran index to represent the spatial auto-correlation that models whether similar documents are retrieved in similar ranks. Higher correlation denotes higher weights of the retriever.

    % \item Ours-post: we use this term to denote the vectorized distance we designed, i.e., $V_{post}(q,R,D)$.

\end{itemize}

% In our parametric combination, we consider Layer Var., Ours-pre, and Ours-post as our component weights for MoR-pre and MoR-post to help forecast the retriever performance from the diverse query only, query-corpus, and query-document semantics.

From Table~\ref{tab:ablation_performance}, we can observe that the signals we designed based on vectorized distances ($V_\text{pre}$ and $V_\text{post}$) are already strong without parametric combination. On the other hand, the classic Moran Index contributes to the parametric combination but is not individually effective in the MoR context. Deep Fusion, similar to our analysis with \textit{Oracle-deep}, improves MoR-post by 3.6\% relatively.

For other signals, intuitive variance-based methods, e.g., Rep. Var., does not show good performance on delegating weights to different retrievers. Further calibration~\cite{zhao2021calibrateuseimprovingfewshot} can be a potential direction to improve this sort of method. However, Thrust and RRF also present to be good signals that can be extracted before and after conducting retrieval. For the concision or the proposed MoR, we did not consider these signals in the parametric combination, yet further performance improvement is anticipated if we do so.

% \clearpage

\subsection{Qualitative Analysis}
\label{sec:qualitative_analysis_appendix}
We qualitatively examine the behavior of \texttt{MoR} in Figure~\ref{tab:retrieval_results}.
The top block illustrates a scenario where most retrievers fail to retrieve the correct passage, yet \texttt{MoR} successfully identifies the relevant one.
The middle block highlights cases where \texttt{MoR} improves retrieval by effectively integrating signals from the most accurate retrievers.
Finally, the bottom block demonstrates that \texttt{MoR}, while generally effective, cannot succeed when all base retrievers fail to retrieve the correct passage.

\definecolor{darkgreen}{rgb}{0,0.5,0}

\begin{table*}[t]
\centering
\small
\begin{tabular}{l l c l}
\toprule
Query & Model & Recall@5 & Top-5 Retrieved IDs \\
\midrule
\multirow{9}{*}{sciq-test\_247}
  & SimCSE       & 0 & train\_7285, test\_58, train\_1445, test\_836, train\_7766 \\
  & ANCE         & 0 & test\_58, train\_2076, train\_7766, train\_8201, train\_8099 \\
  & Contriever   & 0 & train\_7766, train\_4806, train\_8201, validation\_576, train\_7943 \\
  & TASB         & 0 & test\_34, test\_983, train\_8393, train\_4806, train\_3573 \\
  & MPNet        & 0 & train\_2076, train\_7766, train\_8201, test\_836, test\_58 \\
  & GTR          & 0 & train\_2076, test\_836, train\_3573, test\_34, train\_7766 \\
  & DPR          & 0 & train\_2076, train\_8099, test\_58, train\_1944, validation\_576 \\
  & BM25         & 1 & \textcolor{darkgreen}{test\_247}, test\_836, train\_6960, train\_8201, train\_10474 \\
  & \texttt{MoR} & 1 & test\_836, \textcolor{darkgreen}{test\_247}, train\_4806, test\_192, train\_8201 \\
\midrule
\multirow{9}{*}{sciq-test\_0}
  & SimCSE       & 1 & \textcolor{darkgreen}{test\_0}, train\_3722, train\_3328, train\_6009, train\_5098 \\
  & ANCE         & 1 & \textcolor{darkgreen}{test\_0}, train\_6072, train\_6454, train\_4704, train\_6736 \\
  & Contriever   & 1 & \textcolor{darkgreen}{test\_0}, train\_2380, train\_11507, train\_465, train\_7579 \\
  & TASB         & 0 & train\_443, train\_7373, train\_3464, train\_9313, train\_9608 \\
  & MPNet        & 1 & train\_4550, train\_2886, \textcolor{darkgreen}{test\_0}, train\_6009, validation\_355 \\
  & GTR          & 1 & \textcolor{darkgreen}{test\_0}, train\_10570, train\_4550, validation\_355, train\_3299 \\
  & DPR          & 1 & \textcolor{darkgreen}{test\_0}, train\_2837, train\_6072, validation\_582, train\_4485 \\
  & BM25         & 1 & \textcolor{darkgreen}{test\_0}, train\_4704, train\_10381, train\_1544, train\_443 \\
  & \texttt{MoR} & 1 & \textcolor{darkgreen}{test\_0}, train\_4550, train\_11207, test\_696, train\_1997 \\
\midrule
\multirow{9}{*}{sciq-test\_143}
  & SimCSE       & 0 & train\_677, train\_11223, train\_490, train\_2716, train\_5915 \\
  & ANCE         & 0 & train\_7701, train\_5441, train\_2716, validation\_944, train\_877 \\
  & Contriever   & 0 & train\_6186, train\_6460, train\_5114, train\_10363, train\_3753 \\
  & TASB         & 0 & train\_3753, train\_2682, train\_11467, train\_9762, train\_490 \\
  & MPNet        & 0 & train\_2340, test\_669, train\_2716, train\_8417, train\_8583 \\
  & GTR          & 0 & train\_11223, train\_3673, train\_3753, train\_8417, train\_2079 \\
  & DPR          & 0 & train\_7701, train\_5441, train\_2716, train\_877, train\_4268 \\
  & BM25         & 0 & train\_226, train\_3753, train\_11223, train\_2716, train\_3765 \\
  & \texttt{MoR} & 0 & train\_7534, train\_2716, train\_490, train\_998, train\_3753 \\
\bottomrule
\end{tabular}
\caption{Retrieval performance (Recall@5) and top-5 results per model for queries \texttt{sciq-test\_247}, \texttt{sciq-test\_0} and \texttt{sciq-test\_143}, with correct hits highlighted in \textcolor{darkgreen}{dark green}.}
\label{tab:retrieval_results}
\end{table*}

\subsection{Extended Best Retriever Suite Analysis}
\label{sec:extended_best_retrievers_appendix}

\begin{table*}[t]
    \centering
    \small
    \begin{tabular}{l|cccc}
        \toprule
        & \textbf{Best of 2} & \textbf{Best of 3} & \textbf{Best of 4} & \textbf{Best of 5} \\

        \midrule
        Performance & NDCG@5/NDCG@20 & NDCG@5/NDCG@20 & NDCG@5/NDCG@20 & NDCG@5/NDCG@20 \\
        \midrule
        \multirow{3}{*}{Overall} & 56.9/57.9, & 57.2/58.2 & 57.3/58.2 & 57.2/58.1 \\
        & MPNet, GTR & Contriever, GTR, & SimCSE, DPR, & SimCSE, MPNet, \\
        & & MPNet & Contriever, GTR & Contriever, GTR, \\
        & & & & DPR \\
        \midrule
        \multirow{3}{*}{SciQ} & 92.3/92.57, & 92.64/92.77 & 92.77/92.90 & 92.84/92.97 \\
        & Contriever, GTR & Contriever, GTR, & SimCSE, DPR, & SimCSE, DPR, \\
        & & MPNet & Contriever, GTR & Contriever, GTR, \\
        & & & & MPNet \\
        \midrule
        \multirow{3}{*}{SciDocs} & 22.05/29.86 & 22.09/29.60 & 21.89/29.30 & 21.61/28.30 \\
        & SimCSE, MPNet & SimCSE, DPR, & SimCSE, DPR, & SimCSE, DPR, \\
        & & MPNet & ANCE, MPNet & ANCE, Contriever, \\
        & & & & MPNet \\
        \midrule
        \multirow{3}{*}{NFCorpus} & 47.53/40.14 & 47.84/40.34 & 47.90/40.31 & 47.93/40.34 \\
        & Contriever, & ANCE, Contriever, & ANCE, & DPR, ANCE, \\
        & MPNet & MPNet & Contriever, & Contriever, \\
        & & & MPNet, GTR & MPNet, GTR \\
        \midrule
        \multirow{3}{*}{SciFact} & 68.58/72.5 & 68.96/73.08 & 68.43/72.16 & 68.64/72.87 \\
        & Contriever, & Contriever, & SimCSE, ANCE, & DPR, SimCSE, \\
        & MPNet & MPNet, GTR & Contriever, & ANCE, MPNet, \\
        & & & MPNet & Contriever \\
        \bottomrule
    \end{tabular}
    \caption{Best suite of retrievers for \texttt{MoR-post} with different sizes of the retriever list.}
    \label{tab:best_suite_retrievers_appendix}
\end{table*}

In Section~\ref{sec:efficiency}, we discussed the overall best retriever suite across tasks, given a different number of retrievers selected. We also notice that the best suite can be different for different tasks. In this section, we further show the best suite for each task. As shown in Table~\ref{tab:best_retriever_suite}, we can observe that, similar to what we show in the main paper, the best suite of retrievers is not necessarily the best-performing retrievers. 
For example, on SciQ, the top-2 performing retrievers are MPNet and BM25 in terms of NDCG@20. However, the best suite is GTR (supervised) and Contriever (unsupervised). Such observations further validate our intuition in designing \texttt{MoR} - leveraging the comparative advantages among them. Similar findings generalize to other datasets as well.

On the other hand, comparing the retrieval performance of different numbers of \textit{Best of X}, although the overall performance improve consistently with the a larger retriever list, the performance degradation is minor on some tasks. For example, on SciFact, the gap between \textit{Best of 2} and \textit{Best of 5} is 0.06 NDCG@5, which indicate that an efficient version of \texttt{MoR} to be deployed with a curated set of retrievers.

\subsection{Ethical Statements}
We foresee no ethical concerns or potential risks in our work.
All of the retrieval models and datasets are open-sourced, as shown in Table~\ref{tab:dataset_stats} and Section~\ref{sec:appendix_retriever_details}. 
The LLMs we applied in the experiments are also publicly available.
Given our context, the outputs of LLMs are unlikely to contain harmful and dangerous information. The experiments in our paper are mainly on English.

\subsection{Licences of Scientific Artifacts}
We conclude the licenses of the scientific artifacts we used in Table~\ref{tab:tools}. All artifacts are properly used following their original purposes.

\begin{table*} 
    \centering
    \resizebox{\textwidth}{!}{
    \begin{tabular}{llll} 
       \toprule
        \textbf{Artifacts/Packages} & \textbf{Citation} & \textbf{Link} & \textbf{License}\\ 
        \midrule
        SciFact & \cite{wadden-etal-2020-fact} & \url{https://huggingface.co/datasets/BeIR/scifact} & cc-by-sa-4.0 \\
        SciDocs & \cite{cohan-etal-2020-specter} & \url{https://huggingface.co/datasets/BeIR/scidocs} & cc-by-sa-4.0 \\
        SciQ & \cite{welbl-etal-2017-crowdsourcing} & \url{https://huggingface.co/datasets/bigbio/sciq} & cc-by-nc-3.9 \\
        NFCorpus & \cite{boteva2016full} & \url{https://huggingface.co/datasets/BeIR/nfcorpus} & cc-by-sa-4.0 \\
        \midrule
        PyTorch & \cite{paszke-etal-2019-pytorch} & \url{https://pytorch.org/} & BSD-3 License\\
        transformers & \cite{wolf2019huggingface} & \url{https://huggingface.co/transformers/v2.11.0/index.html} & Apache License 2.0\\
        numpy & \cite{DBLP:journals/nature/HarrisMWGVCWTBS20} & \url{https://numpy.org/} & BSD License \\
        matplotlib & \cite{hunter2007matplotlib} & \url{https://matplotlib.org/} & BSD compatible License\\
        vllm & \cite{kwon2023efficient} & \url{https://github.com/vllm-project/vllm} & Apache License 2.0 \\
       \midrule
        LLaMA-3 & \cite{DBLP:journals/corr/abs-2307-09288} & \url{https://huggingface.co/meta-llama/Meta-Llama-3-8B-Instruct} & \href{https://ai.meta.com/llama/license/}{LICENSE}\\
        SimCSE & \cite{gao-etal-2021-simcse} & \url{https://huggingface.co/princeton-nlp/unsup-simcse-bert-base-uncased} & MIT license \\
        Contriever & \cite{izacard2022unsupervised} & \url{https://huggingface.co/facebook/contriever} & \href{https://github.com/facebookresearch/contriever?tab=License-1-ov-file\#readme}{License} \\
        DPR & \cite{karpukhin-etal-2020-dense} & \url{https://huggingface.co/facebook/dpr-ctx_encoder-multiset-base} & cc-by-nc-4.0 \\
        ANCE & \cite{xiong2021approximate} & \url{https://huggingface.co/castorini/ance-dpr-context-multi} & MIT license\\
        TAS-B & \cite{hofstatter2021efficiently} & \url{https://huggingface.co/sentence-transformers/msmarco-distilbert-base-tas-b} & Apache License 2.0\\
        GTR & \cite{ni-etal-2022-large} & \url{https://huggingface.co/sentence-transformers/gtr-t5-base} & Apache License 2.0\\
      \bottomrule
    \end{tabular}}
    \caption{Details of datasets, major packages, and existing models we use. The curated datasets and our code/software are under the MIT License.}
    \label{tab:tools}
\end{table*}

\subsection{Notation}\label{subsec:app:notation}

\begin{table*}[t]
    \centering
    \begin{tabular}{ll}\toprule
        Notation & Description \\\midrule
        % $\instance$ & input instance \\
        % $\inputSpace$ & input space\\
        % $\outputSpace$ & RAG output space\\
        % $\output$ & final RAG output from $\llmModel(\augmentedPrompt)$\\
        % $\target$ & output target \\
        % \\
        % $\queryGenerator(\instance)$ & query generation function\\
        $\query$ & user query\\
        % $\querySpace$ & query space\\
        % \\
        $\corpus$ & corpus (stored retrievable items) \\
        $\doc_j$ & a document in a corpus \\
        \\
        $\rModel_i(\query, \corpus)$ & $i$'th retriever - simply $\rModel_i$ \\
        $L_\rModel$ & set of retrievers \\
        $N$ & the number of retrievers ($|L_\rModel|$)\\
        % $\ranking_i$ & ranked list returned by $\rModel_i$\\
        % $\retrievalScores_i\in\slReals^k$ & retrieval scores returned by $\rModel_i$ \\
        $s_i(\query, \doc_j)$ & query-document relevance score from the $i$'th retriever \\
        \\
        $\MoRModel(\query, \rModel_i, \corpus)$ & \mor weight allocation function \\
        $\tilde{s}(\query, \doc_j)$ & adjusted query-document relevance score after \mor weight aggregation \\
        % $\MoRModelParam$ & \texttt{MoR} model parameter (use: $\MoRModel_\MoRModelParam$)\\
        % $\MoROracleParam$ & Oracle \texttt{MoR} model parameter (Oracle \texttt{MoR}: $\MoRModel_{\MoROracleParam}$)\\
        % $\fusedRanking$ & fused final ranking after \texttt{MoR}\\
        % \\
        % $\promptGenerator(\instance, \fusedRanking)$ & prompt generation function\\        
        % $\augmentedPrompt$ & augmented prompt returned by $\promptGenerator(\instance, \fusedRanking)$\\
        % $\llmModel(\augmentedPrompt)$ & language model \\
        % \\
        % $\utilityMetric(\target, \hat{\target})$ & string utility metric (e.g., EM, F1)\\
        \bottomrule
    \end{tabular}
    \caption{Notation.}
    \label{tab:notations}
\end{table*}

We present a list of the notations we used in Table~\ref{tab:notations} for reference.

\end{document}